\documentclass{ws-procs975x65}

\begin{document}

\title{\uppercase{A new era in data analysis of the cosmological large-scale structure}}

\author{\uppercase{Francisco-Shu Kitaura}}

\address{Leibniz Institute for Astrophysics in Potsdam (AIP), An der Sternwarte 16, 14482 Potsdam\\
E-mail: kitaura@aip.de\\
Schwarzschild fellow}

\begin{abstract}
We report on new developments in the field of large-scale structure analysis. In particular, we present the KIGEN-approach, which allows for the first time for  self-consistent  phase-space reconstructions  from galaxy redshift data.
\end{abstract}

\keywords{Cosmology: large-scale structure of universe; Dark Matter; Dark Energy; Cosmic Web; Galaxy surveys; Data analysis}

\bodymatter

\section{Goal}

What is the underlying nonlinear matter distribution and dynamics at any cosmic time corresponding to a set of observed galaxies in redshift space? \\
To answer this question one needs to recover the  primordial density fluctuations, as an accurate reconstruction of the initial conditions encodes the full phase-space information at any later cosmic time (given a particular structure formation model and a set of cosmological parameters).

\section{Methods}

Previous pioneering attempts to recover the initial conditions have in most of the cases either ignored the relative movement of structures due to gravitation (see e.~g.~Refs.~\refcite{weinberg92},\refcite{kravtsov02},\refcite{klypin03}) or relied on linear theory (Refs.~\refcite{nusser92},\refcite{kolatt96},\refcite{mathis02},\refcite{eisenstein07},\refcite{padmanabhan12},\refcite{doumler13}). Some nonlinear attempts can be found in the literature (see e.~g.~Refs.~\refcite{gramann93},\refcite{croft97},\refcite{narayanan98},\refcite{monaco99},\refcite{kitaura12},\refcite{kitauraetal12a}). Other complex approaches have solved the boundary problem of finding the initial positions of a set of matter tracers governed by the Eulerian equation of motion and gravity with the least action principle (see Refs.~\refcite{peebles89,nusser00,branchini02}). A similar approach consists on relating the observed positions of galaxies in a geometrical way to a homogeneous distribution by minimizing a cost function (Refs.~\refcite{frisch02,brenier03,lavaux10}).  All these approaches have one fundamental aspect in common: they aim at finding a single {\it optimal} solution to the initial conditions boundary problem.

\begin{figure}[h]
  \parbox{0.48\textwidth}{\caption{Flowchart of the KIGEN-code based on a Bayesian Networks Machine Learning approach. The primordial density fluctuations (initial conditions) are obtained from iteratively sampling Gaussian fields, which lead to cosmic structures compatible with the distribution of galaxies given a particular structure formation model. Starting from some initial guess (steps 0-3) structure formation is simulated foward (steps 4-5), observational effects like redshift-space distortions caused by peculiar motions (steps 6-7) and selection function effects according to the magnitude limited survey (step 8) are taken into account and the resulting mock observations (step 9) are matched with the observations in a likelihood comparison process (steps 0-1). The results are used to improve the initial conditions in the next iteration.}}
  \hskip4mm 
\parbox{0.48\textwidth}{\psfig{file=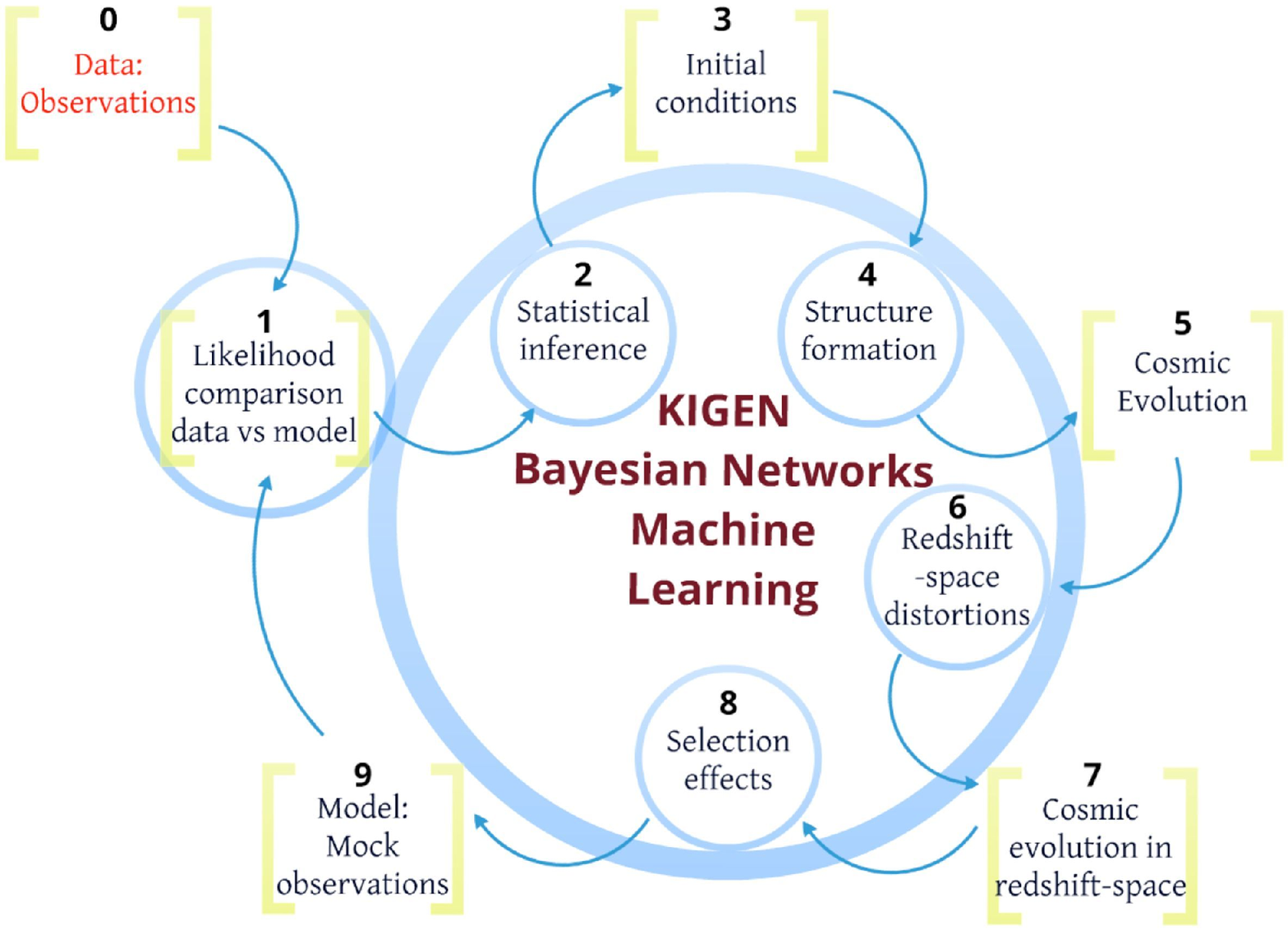, width=0.5\textwidth}}
\label{fig0}
\end{figure}

However, once shell-crossing occurs, two matter tracers can have extremely close positions, but very different peculiar motions and it becomes imposible to know where the tracers came from in a unique way. Moreover, matter collapses to compact objects, which did not exist in the past. Therefore, statistical forward approaches have been introduced (see the KIGEN-code: Ref.~\refcite{kitaura13,kitauraetal12b} and other approaches: Refs.~\refcite{jasche13,wang13}). 
In particular the KIGEN-code permits one for the first time to deal with any kind of structure formation model in a probabilistic way and hereby directly find the ensemble of primordial density fluctuations, which are compatible with the data and such a model. The KIGEN method includes redshift-space distortions (coherent and virialised peculiar motions) in the likelihood comparison to the observations.  In this sense, it is also the first self-consistent phase-space reconstruction method. 

The flowchart of the KIGEN Bayesian Networks Machine Learning approach is presented in Fig.~1.  We note that the likelihood comparison can be used to improve the initial conditions since we have the full information available about the trajectories of the matter tracers from some starting high redshift until the redshift of the observations.

The KIGEN-code has been tested with a semi-analytic halo-model based galaxy mock catalog to demonstrate that the recovered initial conditions are closely unbiased with respect to the actual ones from the corresponding N-body simulation (see Ref.~\refcite{kitaura13}). It has also been applied to the recently released Two-Micron All-Sky Redshift Survey (2MRS: Ref.~\refcite{huchra12}) to perform a cosmography analysis and determine the proper motion of the Local Group finding a close agreement with the direction of the Cosmic Microwave Background  (CMB)  dipole and explaining about 80 \% of its speed  (see Fig.~2 and Ref.~\refcite{kitauraetal12b}), and  to search for the missing baryons in the warm hot inter-galactic medium (see Ref.~\refcite{suarez13}). A thorough analysis of the high performance of the KIGEN-code  and its robustness with constrained N-body simulations has also been done (see Ref.~\refcite{hess13}).

\begin{figure}[h]
\begin{center}
\psfig{file=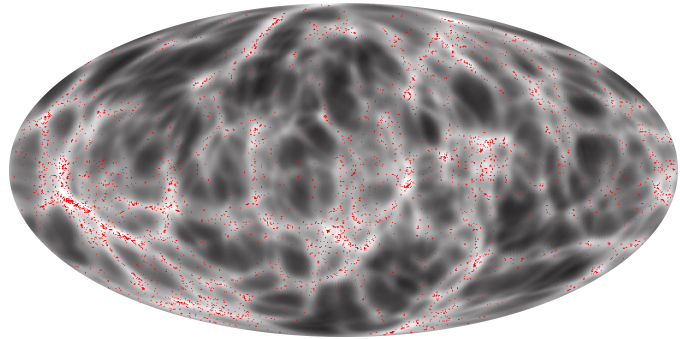, width=.7\textwidth}
\end{center}
\caption{The above figure shows the sky projection of all galaxies in the 2MRS catalog (red dots) at distances of 170 to 280 million light-years and their exquisite correlation with the mean over 25 reconstructed samples of the nonlinear cosmic web (grey scale) using the KIGEN code.}
\label{fig0}
\end{figure}

\section{Conclusions}

We have presented the KIGEN-code: an approach to reconstruct the large-scale structure in the nonlinear regime. We reported on a number of applications of this method. Despite of the flexibility of the KIGEN method, sampling the joint probability distribution function of cosmic density and peculiar velocity fields cannot be massively done with full gravity solvers. Nevertheless,  accurate and efficient structure formation models have been recently developed based on augmenting Lagrangian perturbation theory (ALPT: Ref.~\refcite{kitaurahess13}) and their combination with statistical biasing descriptions  leads to an accurate modeling of the halo distribution (see PATCHY-code: Ref.~\refcite{kitaura14}).
We anticipate a large number of applications based on the approach presented here, ranging from baryon acoustic oscillations and redshift space distortions reconstructions, over environmental studies of different galaxy types, to topological studies of the large-scale structure and measurements of secondary anisotropy signals from the CMB.

\bibliographystyle{ws-procs975x65}

\end{document}